\documentclass[12pt,number,sort&compress]{elsarticle-published}

\biboptions{sort&compress}

\usepackage[psamsfonts]{amssymb}

\usepackage{amssymb}

\usepackage{bm}

\usepackage{siunitx}

\journal{Nucl. Instr. and Meth. A}

\usepackage{hyperref}
\usepackage{comment}

\newcommand{\gsim}{\hbox{ \raise3pt\hbox to 0pt{$>$}\raise-3pt\hbox{$\sim$} }}
\newcommand{\lsim}{\hbox{ \raise3pt\hbox to 0pt{$<$}\raise-3pt\hbox{$\sim$} }}
\newcommand{\del}{\ifmmode{\nabla}         \else{$\nabla$ }               \fi}

\newcommand{\figdir}{./}

\usepackage{graphicx}

\begin{document}

\begin{frontmatter}
\title
{
An Estimation of the Effective Number of Electrons 
Contributing to the Coordinate Measurement with a TPC
}
\author{Makoto Kobayashi}
\address
{
High Energy Accelerator Research Organization (KEK), \\
Tsukuba, 305-0801, Japan
} 

\begin{abstract}

For time projection chambers (TPCs) the accuracy in measurement of 
track coordinates along the pad row direction deteriorates with the
drift distance ($z$):
$\sigma_{\rm X}^2 \sim D_{\rm X}^2 \cdot z / N_{\rm eff}$, where
$D_{\rm X}$ is the diffusion constant and $N_{\rm eff}$ is the
effective number of electrons.
Experimentally it has been shown that $N_{\rm eff}$ is smaller than
the average number of drift electrons per pad row ($\overline{N}$).
We estimated $N_{\rm eff}$ by means of a simple numerical simulation.
The simulation shows that $N_{\rm eff}$ can be as small as $\sim$ 30\% 
of $\overline{N}$ due to the combined effect of statistical
fluctuations in the number of drift electrons ($N$) and in their
multiplication in avalanches.

\end{abstract}


\end{frontmatter}


\section{Introduction}

The spatial resolution of TPCs along the pad row direction is
expressed as
\begin{displaymath}
\sigma_{\rm X}^2 = \sigma_{0 \rm X}^2 + D_{\rm X}^2 \cdot z\;,
\end{displaymath}
where $\sigma_{0 \rm X}$ is the intrinsic resolution, $D_{\rm X}$ takes
into account the diffusion, and $z$ is the drift distance.
On the other hand the width of pad response is given by
\begin{displaymath}
\sigma_{\rm PR}^2 = \sigma_{0 \rm PR}^2 + D_{\rm T}^2 \cdot z\;,
\end{displaymath}
with $D_{\rm T}$ being the transverse diffusion constant\footnote
{In presence of axial magnetic field ($B$) the transverse diffusion
constant is given by $D_{\rm T}(B=0)/\sqrt{1+\omega^2\tau^2}$,
where $\omega \equiv eB/m$,
the electron cyclotron frequency and $\tau$ is the mean free time of
drift electrons.  
$D_{\rm T}$ is related to the diffusion coefficient ($D$) thorough
$D_{\rm T}^2 = 2D/W$, where $W$ is the
electron drift velocity.  
}.
Let us forget about $\sigma_{0 \rm X}$ and $\sigma_{0 \rm PR}$ here
and focus on
the $z$-dependent (diffusion) terms although the origins of
the intrinsic terms themselves are of great interest.

$D_{\rm X}^2$ is expected to be proportional to $D_{\rm T}^2$ and
expressed as
\begin{displaymath}
D_{\rm X}^2 = \frac{D_{\rm T}^2}{N_{\rm eff}}\;,
\end{displaymath}
where $N_{\rm eff}$ is the effective number of electrons.
One may naively expect $N_{\rm eff}$ to be the average number of drift
electrons per pad row ($\overline{N}$).
In reality, however, $N_{\rm eff}$ is significantly smaller than
$\overline{N}$~\cite{Carnegie,Ikematsu,Colas}:
\begin{displaymath}
N_{\rm eff} = \frac{\overline{N}}{R}\;,
\end{displaymath}
where $R$ is the reduction factor ($> 1$).

In what follows an attempt is made to estimate $N_{\rm eff}$
(or equivalently, $R$) for a typical TPC operated in argon-based gases,
by means of a numerical simulation.

\section{Expectations \label{Expectations}}

Let us assume:
\begin {itemize}
\item Readout pads are aligned along the $x$ axis and charged particles
      traverse the drift volume of the TPC, in parallel with the
      readout plane ($z$ = constant) and in perpendicular to the
      pad rows ($x$ = constant = $x_0$).
\item A charged particle leaves $N$ electrons along its path, to be
      detected by a single pad row. 
      The initial electron clusters are considered as point-like.
      The $x$ coordinate of each electron then deviates from
      $x_0$ during the drift towards a detection gap\footnote
{
      The detection gap may be equipped with 
      a multi-wire proportional chamber (MWPC) or
      a micro-pattern gas detector (MPGD), along with readout pads.
}
      because of transverse diffusion.
      The standard deviation along the pad row direction at the
      entrance of the detection gap is denoted by $\sigma_x
      (\equiv D_{\rm T}\sqrt{z})$\footnote
{
      Let us ignore 
      displacement of the coordinates caused by
      the {\boldmath $E$} $\times$ {\boldmath $B$} effect
      near the entrance of the detection gap
      and possible ``quantization effects'' on the coordinates due to
      finite granularity of amplifying elements in the case of
      MPGD readout.
}.
\item Each drift electron gets amplified in the detection gap and
      the charge ($q$) is collected by (induced on) the pads.
      The charge multiplication process (avalanche)
      of each electron develops independently 
      of those initiated by other electrons and the fluctuation of
      avalanche size for the $i$-th electron ($1 \leq i \leq N$) is
      given by the Polya distribution~\cite{Alkhazov}:
\begin{displaymath}
      P(q_i) = \frac{1}{\overline{q}} \cdot 
      \frac{(1 + \theta)^{1 + \theta}}{\Gamma (1 + \theta)} \cdot
      \Bigl( \frac{q_i}{\overline{q}} \Bigr)^{\theta} \cdot
      {\rm exp} \displaystyle \Bigl( -(1 + \theta)
      \frac{q_i}{\overline{q}} \Bigr)\;,
\end{displaymath}
      where $\overline{q}$ is the average charge and $\theta$ is a
      free parameter determining the shape of distribution.
\item The width of the pads is small enough so that the center of
      gravity of charge distribution on the pads caused by the $i$-th
      electron be equal to its arrival position $x_i$ at the entrance
      of the detection gap.
\item The ``measured'' $x$ coordinate of the track ($X$) is
      defined as the charge
      centroid of the pad response, which is the superposition of the
      contribution from each of the drift electrons.
\end{itemize}

Under these assumptions and definitions the track coordinate measured
with the pad row is given by a weighted mean of $x_i$:
\begin{displaymath}
     X = \frac{\sum\limits_{i = 1}^N q_i \cdot x_i}
               {\sum\limits_{i = 1}^N q_i}\;.
\end{displaymath}

The expected spatial resolution is readily obtained as follows\footnote
{
From now on, both angle brackets and an over-line denote the
average of the quantity in-between or beneath.
}.
\begin{eqnarray*}
\nonumber
\sigma_X^2 &=& \left< (X - x_0)^2 \right> \\
     &=& \left< \left( \frac{\sum_{i=1}^N q_i \cdot (x_i - x_0)}
                              {\sum_{i=1}^N q_i}  \right)^2 \right> \\
     &=& \left< \frac{1}{\left(\sum_i q_i \right)^2}
           \left( \sum\limits_i q_i^2 \cdot (x_i-x_0)^2
                \;\;+\;\; \sum_{i \not\neq j} q_i \cdot q_j 
                      \cdot (x_i-x_0) \cdot (x_j-x_0) \right) \right> \\
     &=& \left< \frac{1}{\left(\sum_i q_i \right)^2}
             \left( \left< (x-x_0)^2 \right>_x \cdot \sum\limits_i q_i^2
                \;\;+\;\; \left< x-x_0 \right>_x^{\;2}
                         \cdot \sum_{i \not\neq j} q_i \cdot q_j
                                                    \right) \right>_q \\
     &=& \left< \frac{\sum_i q_i^2}{\left(\sum_i q_i \right)^2} \right>
                                       \cdot \left< (x-x_0)^2 \right> \\
     &=& \left< \frac{\sum_i q_i^2}{\left(\sum_i q_i \right)^2} \right>
                                                    \cdot \sigma_x^2 \; ,
\end{eqnarray*}
where the symbol $\left< \; \cdot \cdot \cdot \; \right>_{x\;(q)}$
stands for the average taken over a variable (variables) $x\;(q_i)$. 
Hereafter we assume 
$Q \equiv \sum\nolimits_{i=1}^N q_i = {\rm constant} = N \cdot
\overline{q}$, expecting $N \gg 1$.
Then
\begin{displaymath}
\sigma_{\rm X}^2 = \frac{\sigma_x^2}{N} \cdot
                      \frac{\overline{q^2}}{\overline{q}^2}
                         \equiv \frac{\sigma_x^2}{N} \cdot (1 + f)\;,         
\end{displaymath}
with $f \equiv \sigma_{\rm q}^2 / \overline{q}^2$,
the relative variance of $q$.

In fact, $N$ is not a constant and fluctuates according to its
probability density function $P(N)$.
The expected variance in this case is given by
\begin{eqnarray}
\sigma_{\rm X}^2 &=& \sum\limits_{N=1}^{\infty} P(N) \cdot
                       \frac{\sigma_x^2}{N} \cdot (1 + f)  \nonumber \\
                 &=& \sigma_x^2 \cdot (1+f) \cdot
                       \sum\limits_{N=1}^{\infty} P(N) \cdot
                            \frac{1}{N}  \nonumber \\
                 &=& \sigma_x^2 \cdot (1+f) \cdot
                   \displaystyle \bigl ( \overline{\frac{1}{N}} \bigr )
                                                           \nonumber \\
                 &=& \frac{\sigma_x^2}{\overline{N}} \cdot (1+f) \cdot
                        \overline{N} \cdot
                   \displaystyle \bigl ( \overline{\frac{1}{N}} \bigr )
                                                           \nonumber \\
                 &\equiv& \frac{\sigma_x^2}{N_{\rm eff}}
               = \frac{D_{\rm T}^2}{N_{\rm eff}} \cdot z\;,\label{eq:1}
\end{eqnarray}
where $N_{\rm eff} \equiv \overline{N}/R$, with
$R \equiv (1+f) \cdot \overline{N} \cdot
       \overline{N^{-1}}$.
It is worth noting that one needs to evaluate $\overline{N^{-1}}$
instead of $\overline{N}$ for the estimation of $N_{\rm eff}$. 
The reduction factor ($R$) is a product of statistically independent
quantities: $R = R_{\rm q} \cdot R_{\rm N}$, where
$R_{\rm q} \equiv 1+f$ and $R_{\rm N} \equiv \overline{N}
                                        \cdot \overline{N^{-1}}$.
The relative variance of Polya distribution is given by
$f = 1/(1+\theta)$.
If we assume $\theta = 0.5$, the frequently used value in wire-chamber
simulations,
$f = 2/3$, yielding $\sim$ 1.67 for $R_{\rm q}$\footnote
{
When $\theta = 0$ the Polya function gives the Furry (exponential)
distribution, which is sometimes used to describe avalanche
fluctuations.
In that case, $R_{\rm q}$ is expected to be $\sim$ 2. 
}.

It should be noted, however, that $Q \equiv \sum\nolimits_{i=1}^N q_i$
has been assumed to be constant for a given number $N$.
The validity of this approximation is checked by a numerical simulation
(see the next section).
As for the estimation of $R_{\rm N}$ we rely totally on the simulation.

\section{Numerical simulation}

\subsection{Assumptions}

The simulation was carried out taking into account the following:
\begin{enumerate}
\item {\bf Primary ionization statistics} (Poissonian) \\
      The mean number of primary ionization acts per pad row
      ($\overline{N_{\rm P}}$)
      is assumed to be $N_{\rm CL}$ $\times$ 0.63 cm $\times$ 1.2,
      where $N_{\rm CL}$ is the average primary ionization density
      for minimum ionizing particles [cm$^{-1}$],
      0.63 cm is the pad row pitch
      ($\gsim$ pad length)
      and the factor 1.2 takes account of the relativistic rise
      for 4 GeV/c pions
      (see, for example, fig. 8 of Ref.~\cite{Lapique})\footnote
{
      The pad row pitch and the incident particles chosen here are
    typical of those in the experiments of Refs.~\cite{Ikematsu,Colas}.
}.     
      We use 29.4 cm$^{-1}$~\cite{Sauli} or
      24.3 cm$^{-1}$~\cite{Sharma} for the value of $N_{\rm CL}$
      in argon.
\item {\bf Cluster size distribution} \\
      We use the probability density listed in table 2 of
      Ref.~\cite{Fischle} for argon (column (b)).
      This distribution, combined with the primary ionizaion statistics
      mentioned above, generates the probability density function
      $P(N)$ in Eq. (\ref{eq:1}).
\item {\bf Diffusion}: fluctuation of $x_i$ (Gaussian) \\
      The {\em standard\/} normal distribution is chosen arbitrarily
      just for simplicity.
      This defines a dimensionless transverse distance scale.
\item {\bf Avalanche fluctuation} \\
      The Polya type with $\theta$ = 0.5.
      $\overline{q}$ is set arbitrarily to 1.
\end{enumerate}
See Figs. 1--3 for the fundamental distributions.
It should be noted that the chamber gas is assumed to be
pure argon at NTP.

\begin{figure}[http]
    \begin{tabular}{ccc}
      \begin{minipage}[t]{0.30\hsize}
        \centering
        \includegraphics*[scale=0.30]{\figdir/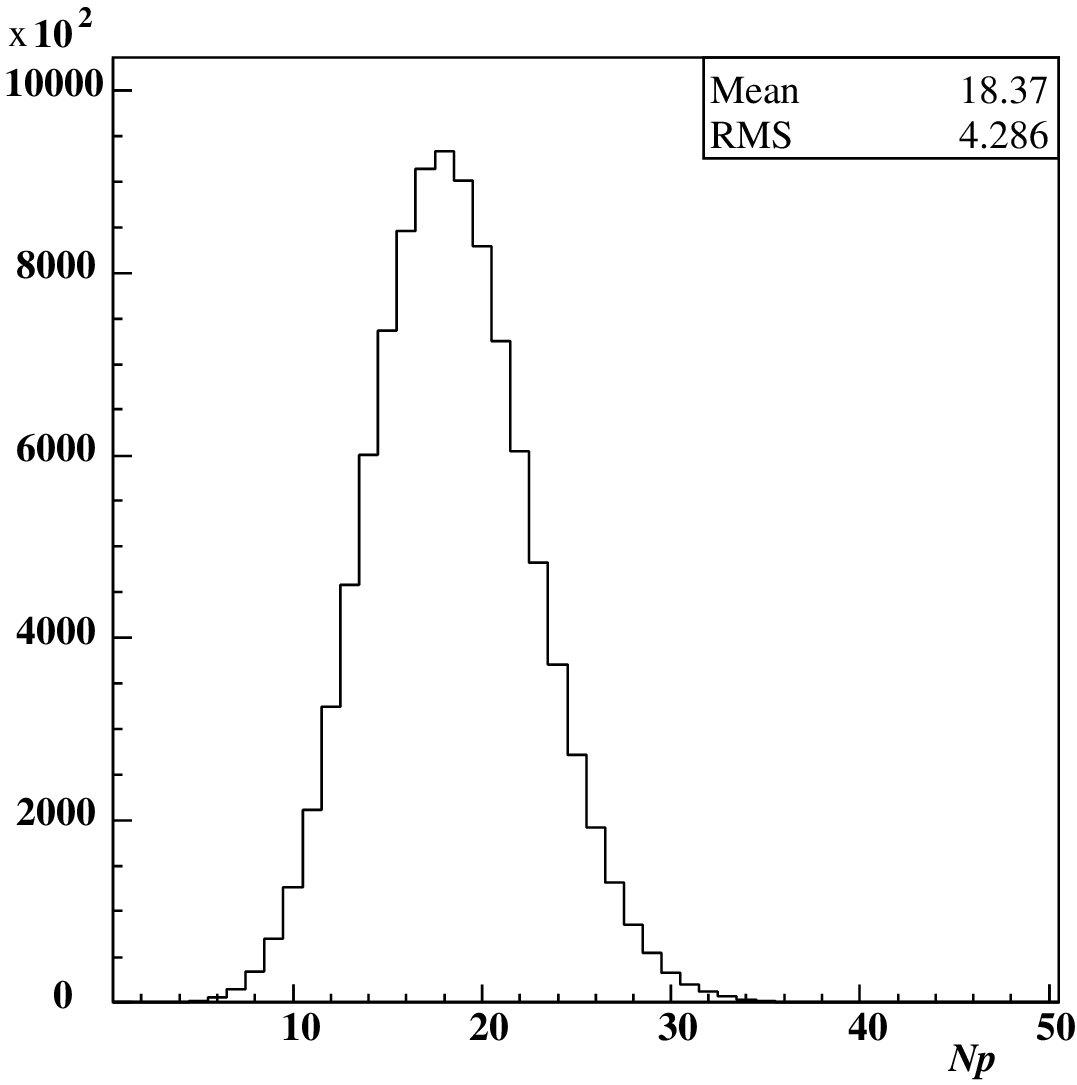}
        \caption{\footnotesize Distribution of the number of primary ionizations
              ($N_{\rm P}$). $N_{\rm CL}$ is assumed to be 
                                                   24.3 cm$^{-1}$.}
        \label{fig1}
      \end{minipage} &
      \begin{minipage}[t]{0.30\hsize}
        \centering
        \includegraphics*[scale=0.30]{\figdir/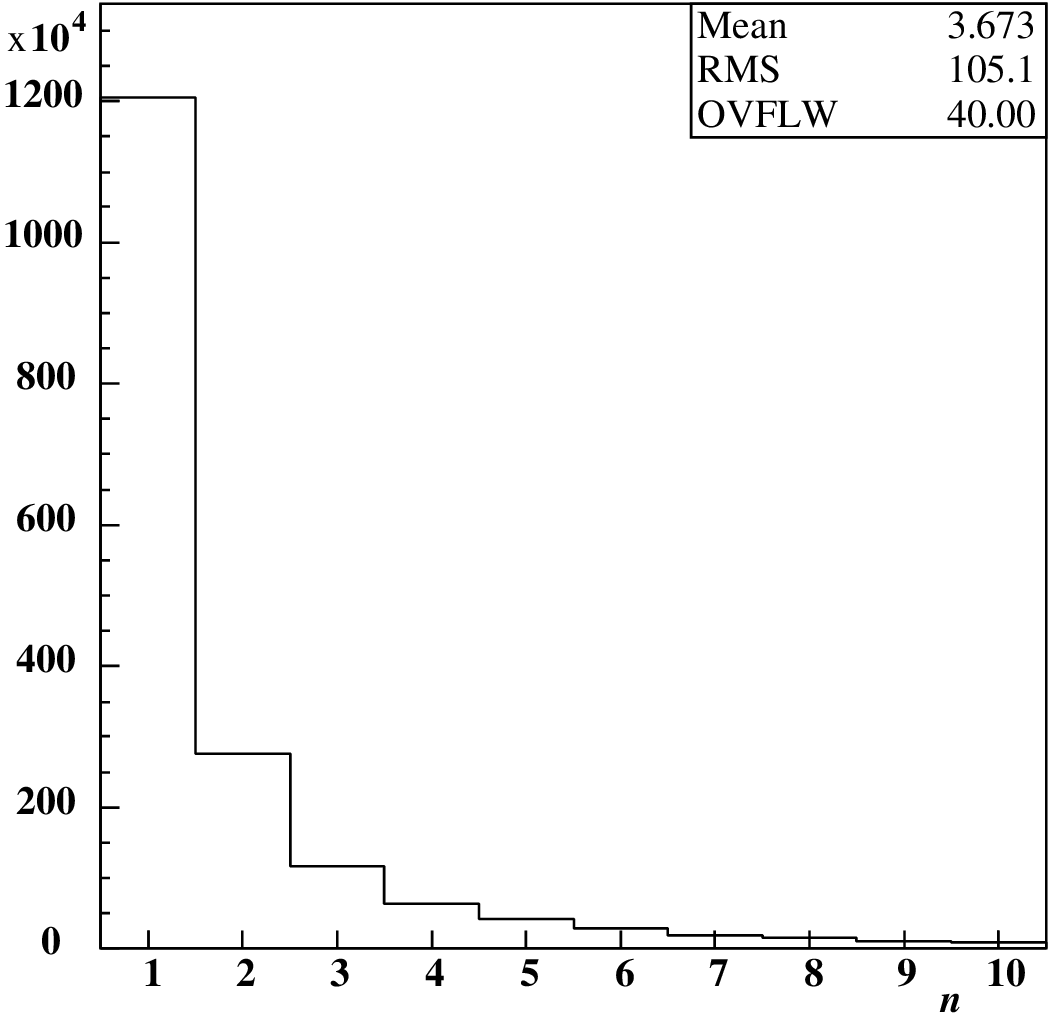}
        \caption{\footnotesize Distribution of the cluster size ($n$).}
        \label{fig2}
      \end{minipage} &
      \begin{minipage}[t]{0.30\hsize}
        \centering
        \includegraphics*[scale=0.30]{\figdir/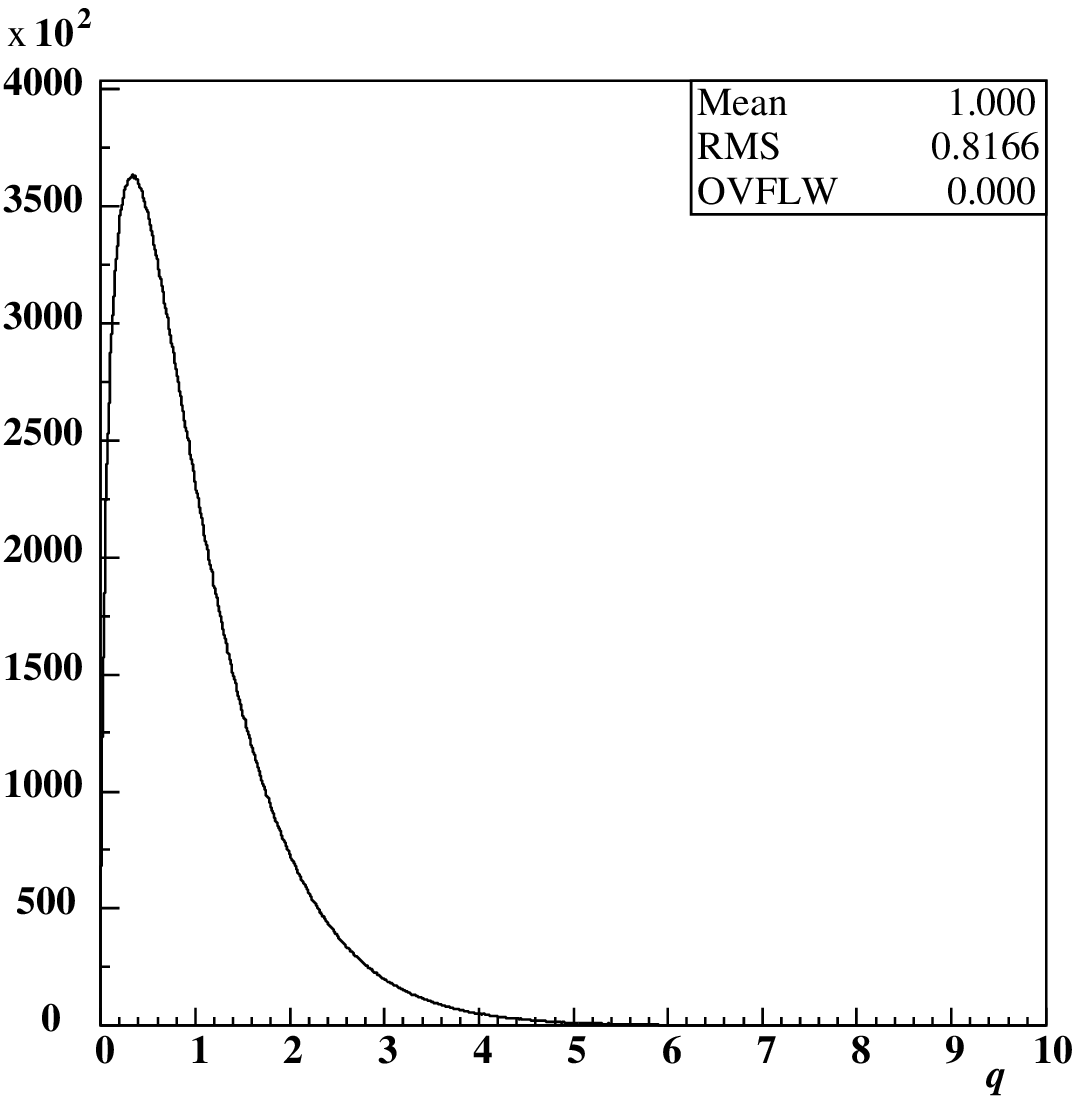}
        \caption{\footnotesize Distribution of the avalanche charge ($q$):
            the Polya type with $\theta = 0.5$ and $\overline{q} = 1$.}
        \label{fig3}
      \end{minipage}
    \end{tabular}
\end{figure}

\subsection{Results}

Table 1 summarizes the results given by the simulation.
See also Figs. 4--6 for the relevant distributions.
The $x$ coordinate of particle tracks ($x_0$) is arbitrarily chosen
to be 0 in Fig. 4.
It should also be recalled that $\sigma_x$ is set to 1.0 in the
simulation.
%
\renewcommand{\arraystretch}{2.0}
\begin{center}


Table 1. Summary of the simulation results.

\bigskip

\begin{tabular}{|l||c|c|c|c|} \hline

 Symbol & Definition & \hspace{3mm}Value\hspace{3mm} &
                              Remarks & Related Figs.
                                               \\ \hline \hline
   $N_{\rm eff}$ & $\displaystyle 
                   \frac {\sigma_x^2}{\sigma_{\rm X}^2}$  &
                                       27.7  &  & $-$ \\ [1mm]
                                              \cline {1-3} \cline{5-5}
   $R$ & $\displaystyle \overline{N}
      \cdot \frac {\sigma_{\rm X}^2}{\sigma_x^2}$ & 2.57 &
                      $N_{\rm CL}$ = 29.4 cm$^{-1}$  & $-$  \\ [1mm]
                                              \cline {1-3} \cline{5-5}
   $R_{\rm N}$ & $\overline{N} \cdot
       \displaystyle \bigl (\overline{\frac{1}{N}} \bigr )$ &
     1.59  & $\overline{N} = 71$ & $-$ \\ [1mm]
                                              \cline {1-3} \cline{5-5}
   $R_{\rm q}$  & $\displaystyle \frac {R}{R_{\rm N}}$ &
                                     1.62  &  & $-$     \\ [1mm] 
                                                         \hline \hline
   $N_{\rm eff}$ & $\displaystyle 
                   \frac {\sigma_x^2}{\sigma_{\rm X}^2}$  &
                                      22.2  &  & Fig. 4 \\ [1mm]
                                              \cline {1-3} \cline{5-5}
   $R$ & $\displaystyle \overline{N}
      \cdot \frac {\sigma_{\rm X}^2}{\sigma_x^2}$ & 3.22 & 
                  $N_{\rm CL}$ = 24.3 cm$^{-1}$ & Figs. 4--5 \\ [1mm]
                                              \cline {1-3} \cline{5-5}
   $R_{\rm N}$ & $\overline{N} \cdot
       \displaystyle \bigl (\overline{\frac{1}{N}} \bigr )$ &
     2.00  & $\overline{N} = 71$ & Figs. 5--6 \\ [1mm]
                                              \cline {1-3} \cline{5-5}
   $R_{\rm q}$  & $\displaystyle \frac {R}{R_{\rm N}}$ &
                                       1.61  &  & $-$   \\ [1mm]
                                                                \hline
\end{tabular}  


\end{center}

\begin{figure}[http]
    \begin{tabular}{ccc}
      \begin{minipage}[t]{0.30\hsize}
        \centering
        \includegraphics*[scale=0.33]{\figdir/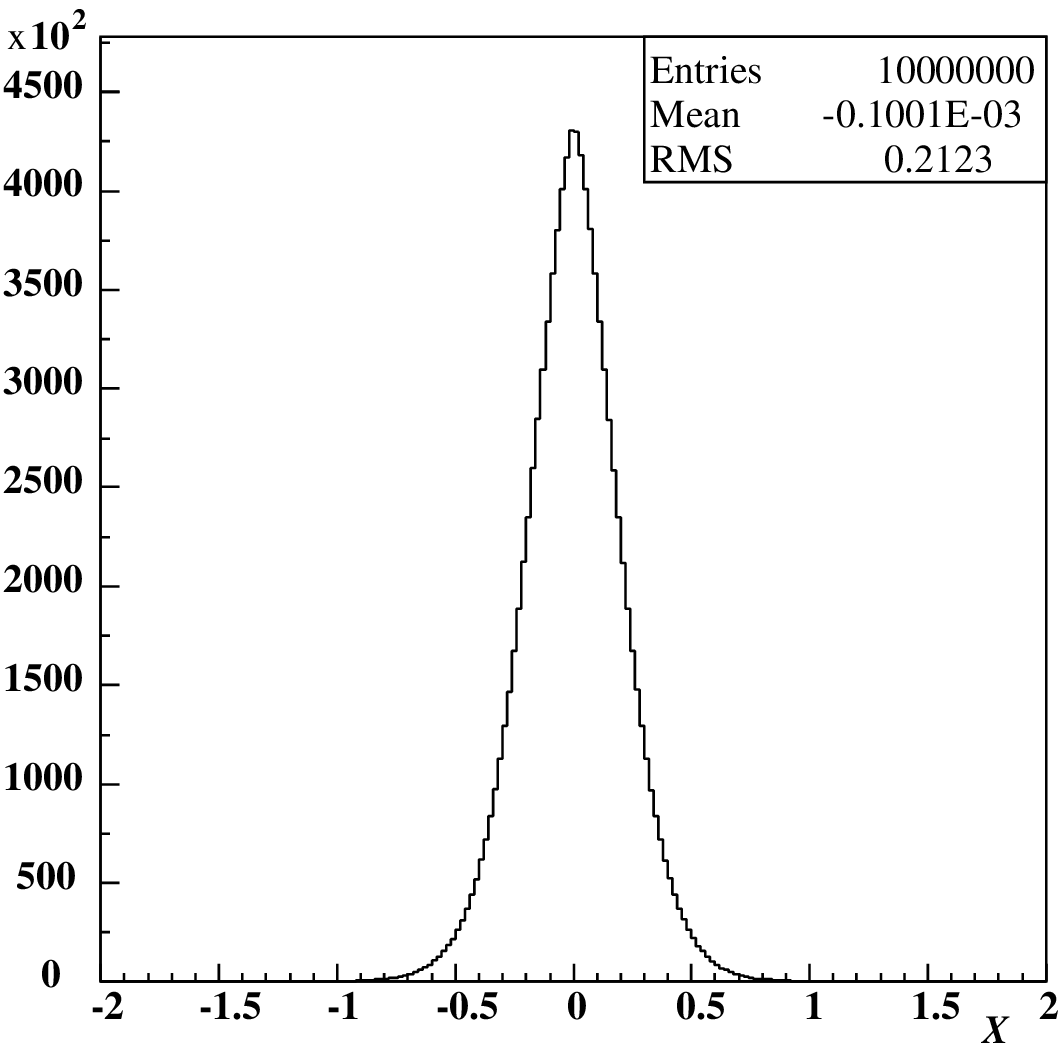}
        \caption{\footnotesize Distribution of the {\em measured\/} 
                                              track coordinates ($X$).}
         \label{fig4}
      \end{minipage} &
      \begin{minipage}[t]{0.30\hsize}
        \centering
        \includegraphics*[scale=0.33]{\figdir/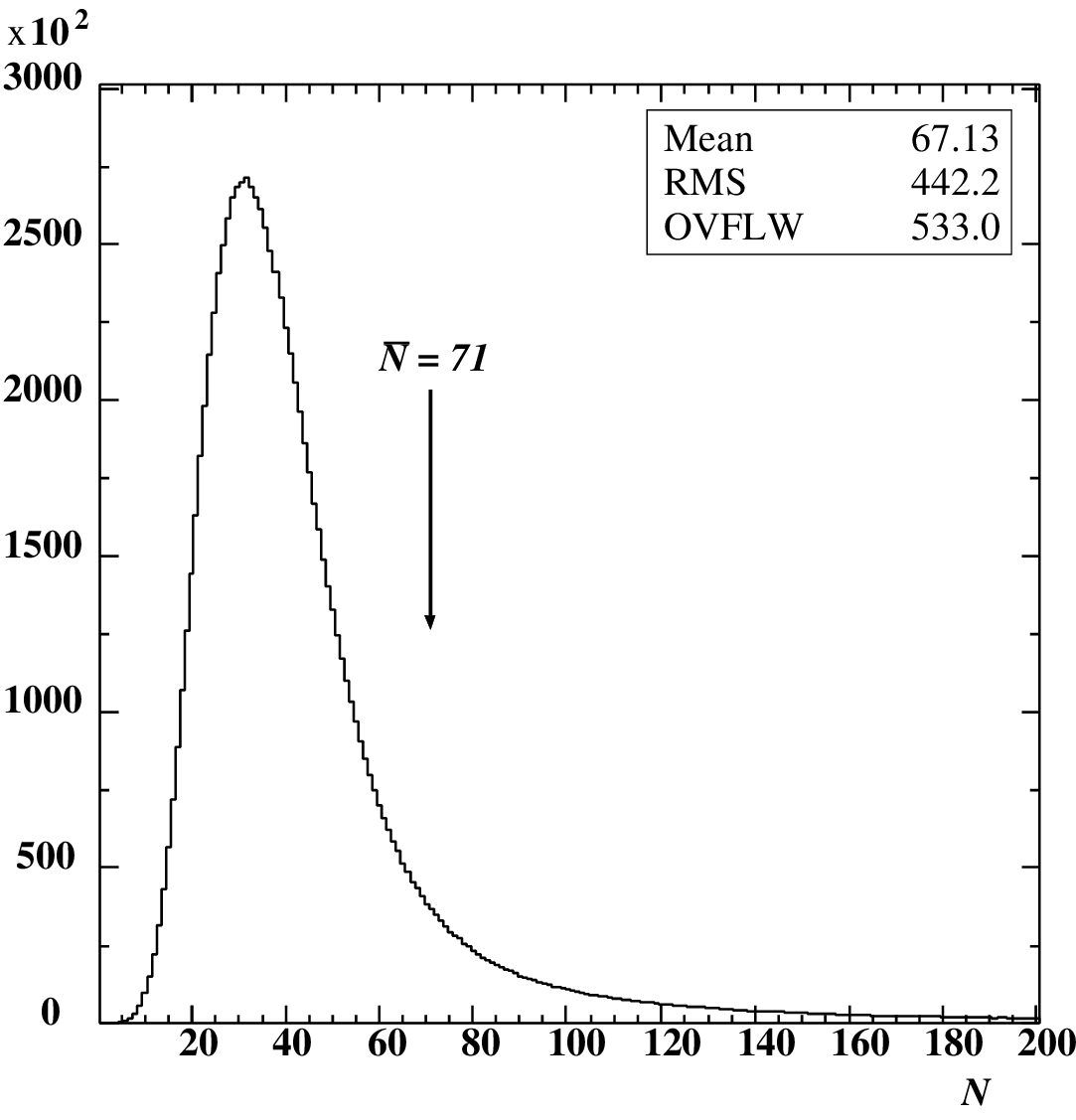}
        \caption{\footnotesize Distribution of the total number of
                                                drift electrons ($N$).}
         \label{fig5}
      \end{minipage} &
      \begin{minipage}[t]{0.30\hsize}
        \centering
        \includegraphics*[scale=0.33]{\figdir/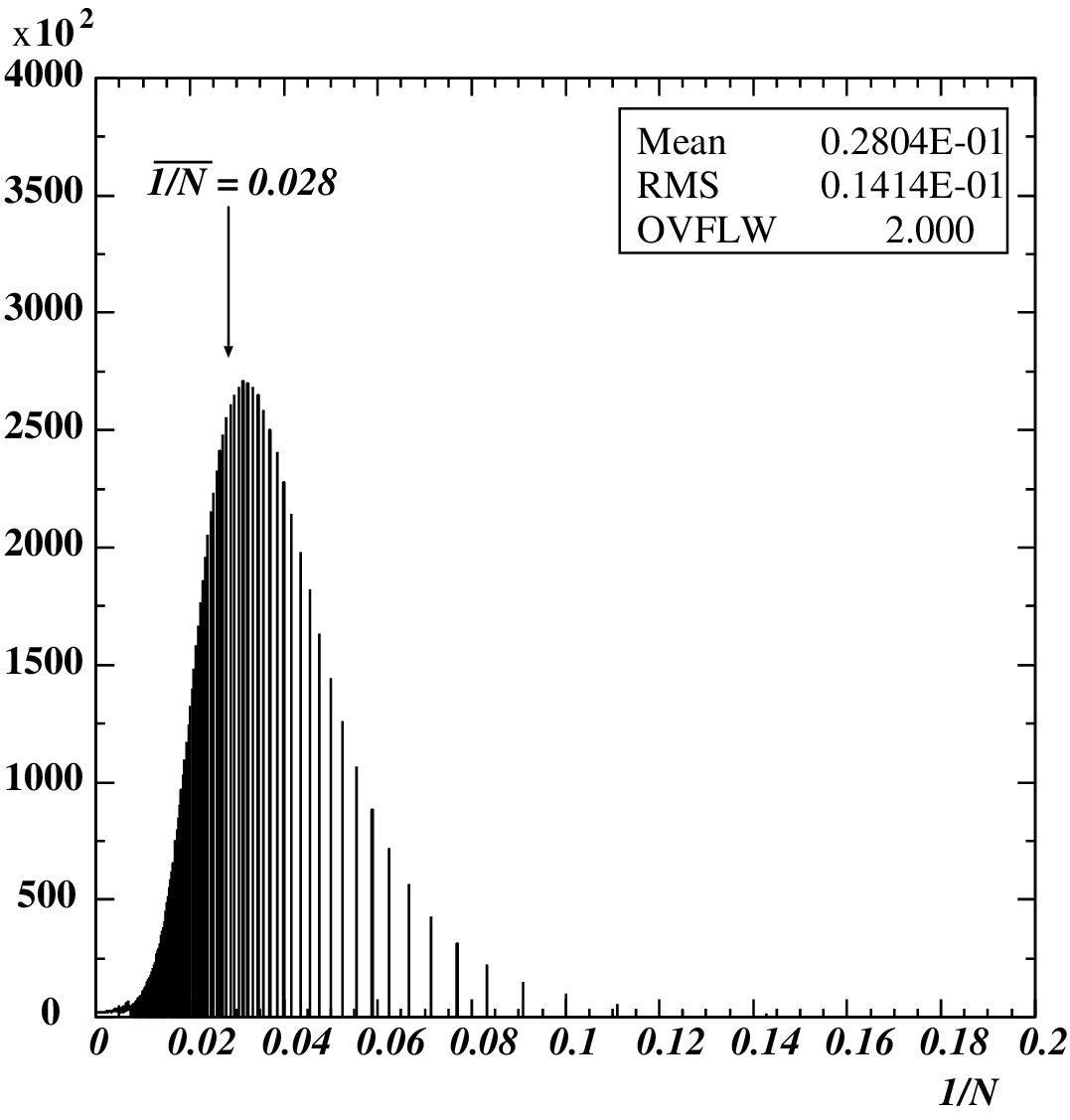}
        \caption{\footnotesize Distribution of $1/N$.}
        \label{fig6}
      \end{minipage}
    \end{tabular}
\end{figure}

The simulation results deserve several comments below:
\renewcommand{\labelitemi}{$\ast$}
\begin{itemize}
\item We are not sure if the Polya distribution represents the
      fluctuation in the avalanche size correctly, especially
      in the case of (cascaded) GEM~\cite{SauliGEM} or
      MicroMEGAS~\cite{Giomataris}.
\item Granting that the Polya type is an appropriate distribution,
      the value of parameter $\theta\/$ should depend on gas mixtures
      and also on the type of detection device: MWPC, GEMs
      or MicroMEGAS,
      and its operating high voltages.
\item It may be a bold assumption to suppose no interference between
      avalanches initiated by different drift electrons when their
      distances are small.
\item The simulation assumes pure argon as the chamber gas and possible
      effects of quenchers, 
      such as increase of $\overline{N}$ due to the Penning effect,
      are not included.
\item The simulation relies on a measurement of cluster size
      distribution~\cite{Fischle}, which is a little bit different from
      the result of a simulation~\cite{Lapique}.
\item In reality, electron clusters created along a particle path
      are not point-like and have finite sizes in space
      (intrinsic track width).
      However, this hardly affects the $z$-dependence of spatial
      resolution ($D_{\rm X}$).
\item Drift electrons experience further diffusion in the
      multiplication region (detection gap).
      However, this does not contribute to $D_{\rm X}$ defined in the
      drift region either.
\item The simulation assumes ideal readout electronics,
      free from noise, and
      with a dynamic range wide enough to record tracks containing
      large cluster(s) and/or large avalanche(s) correctly without
      saturation of amplifiers or ADC-count overflows.
\item $R_{\rm q}$ is (approximately) given once the value of
      $\theta\/$ is fixed.
      However, 
      $R_{\rm N} (\equiv  \overline{N} \cdot \overline{N^{-1}}$)
      is sensitive to the cut on the maximum cluster size
      ($n_{\rm max}$)\footnote
{
      The cluster size distribution has a long tail in proportion to
      $1/n^2$, with $n$ the cluster size.
      Therefore a cut has to be applied for large $n$ in order to keep
      the average total number of electrons ($\overline{N}$) finite.
      In the simulation $n_{\rm max}$ is set somewhat arbitrarily
      to reproduce $\overline{N}$ consistent with the total ionization
      density given in Refs.~\cite{Sauli,Sharma}
      (94 cm$^{-1}$ for minimum ionizing particles).
}
      since $\overline{N}$ inevitably depends on $n_{\rm max}$ while
      $\overline{N^{-1}}$ is fairly constant over a wide range of
      assumed $n_{\rm max}$\footnote
{
      This was confirmed by the simulation with several numbers between
      $10^2$ and $10^6$ for $n_{\rm max}$.
}.
      On the other hand 
      $N_{\rm eff} \sim 1/\{ (1+f) \cdot \overline{N^{-1}} \}$
      is rather stable against variation of $n_{\rm max}$.
      It should be noted, however, that
      $N_{\rm eff}$ as well as $R_{\rm N}$ depends 
      on $\overline{N_{\rm P}}$ ($N_{\rm CL}$) and on the
      probability densities assigned in the populated region of the
      cluster size distribution\footnote
{
      This was confirmed by intentionally manipulating the cluster size
      distribution.
}.
\end{itemize}

\section{Summary}

We may summarize the results as follows:
\renewcommand{\labelitemi}{$\bullet$}
\begin{itemize}
\item We have carried out a simple numerical simulation of
      the spatial resolution given by a TPC
      operated in argon-based gases, 
      taking into account the fluctuations in the total number of
      drift electrons and in their multiplication in avalanches.
      The simulation gives 22--28 for the effective number of
      electrons ($N_{\rm eff}$) contributing to the coordinate
      measurement along the pad row, depending on the assumed primary
      ionization density ($N{\rm_{CL}}$). 
      These numbers correspond to 30--40\% of the 
      average number of drift electrons per pad row made up of
      6-mm-long pads ($\overline{N}$ $\sim$ 71).
\item Contributions of the two fluctuations, i.e. ionization
      statistics and avalanche multiplication, to the reduction factor
      ({\it R\/}) are found to be of comparable size.
\item $R_{\rm q}$ is close to the expected value quoted in
      section~\ref{Expectations} ($\sim$ 1.67),
      justifying, to some extent, the ``constant $Q$'' approximation
      and the resultant Eq. (\ref{eq:1}).
\item In view of the contribution of diffusion to the spatial
      resolution it is advantageous to use a gas with a small
      diffusion constant ($D_{\rm T}$), a large electron yield
      ($\propto \overline{N}$), 
      a high average primary ionization density 
      ($N_{\rm CL}$) and a small avalanche fluctuation ($f$)\footnote
{
      Notice that 
      $\sigma_{\rm X}^2 = \sigma_x^2/N_{\rm eff}$
      with $\sigma_x^2 = D_{\rm T}^2 \cdot z$,
      and that 
      $N_{\rm eff} = \overline{N}/(R_{\rm q} \cdot R_{\rm N})$,
      where 
      $R_{\rm q} \sim 1 + f$
      and $R_{\rm N}$ is 
      a decreasing function of $\overline{N_{\rm p}}$ as is expected. 
      We recall 
      that $f$ should depend on the detection device 
      and its operating conditions as well.
},
      as well as long mean free time ($\tau$) in presence of
      axial magnetic field.
      In this respect longer pads are also advantageous
      in the case of conventional TPCs with discrete pad rows since
      $\overline{N_{\rm P}}$ (as well as $\overline{N}$) is larger,
      giving a smaller $R_{\rm N}$\footnote
{
It should be noted that $R_{\rm N}$ is greater than unity because of
the difference between the mode and the average of $N$ distribution,
which is smaller for larger $\overline{N_{\rm P}}$.
See also Table 1 and compare $R_{\rm N}$ for the different values of
$N_{\rm CL}$.
},
      but at the expense of a larger angular pad effect.
\item An argument similar to that employed in section 2 is applicable
      also to the coordinate measurement in the drift direction
      by substituting $Z$, $z_i$ and $\sigma_z$ respectively for
      $X$, $x_i$ and $\sigma_x$.
\end{itemize}

The presented method for estimation of the effective number of
electrons can be applied for a given gas mixture
and for a given amplifying and readout scheme including pad size
if the relevant parameters are available.

\section*{\nonumber Acknowledgments}

The author would like to thank many colleagues for their continuous 
encouragement and support.
He is especially grateful to Dr. Keisuke Fujii of KEK and 
Dr. Khalil Boudjemline at Carleton university for fruitful discussions.



\end{document}